\input epsf
\documentstyle[prl,aps]{revtex}
\tighten
\begin{document}
\draft

\title{
Novel Circuit Theory of Andreev Reflection
       }

\author{ Yuli V. Nazarov}
\address{
Faculteit van  Technische Natuurwetenschappen and DIMES, 
Technische Universiteit Delft,\\
 Lorentzweg 1,2628 CJ Delft, the Netherlands
	}
\maketitle
\begin{abstract}
We review here a novel circuit theory of
superconductivity. The existed circuit theory of Andreev
reflection has been revised to account for decoherence between
electrons and holes and twofold nature of the distribution function.
The description of arbitrary connectors has been elaborated.
In this way  one can cope with the most of the factors  
that limited applicability of the old circuit theory.
We give a simple example and discuss numerical implementation of the theory. 
\end{abstract} 


\section{Introduction}\label{intro}

Superconductivity is by virtue a mesoscopic phenomenon
since the typical length scales involved exceed by far
Fermi wavelenght and, frequently, the mean free path.\cite{Tinkham}
The adequate semiclassical theory of non-equlibrium 
superconductivity has been initially elaborated for bulk systems.
\cite{LarkOvch1,LarkOvch2}
Subsequent contributions \cite{Blonder,Zaitsev,Kuprianov} 
that treated supeconducting
constrictions and interfaces have provided the extension
of the theory to heterogenious structures and the systems
made artificially. Thereby the theoretical development
has been essentially accomplished. 
The concise account for this work one can find in \cite{Klapwijk}.
We will refer to this patch of theoretical development
as to "full theory".
 
There used to be a sharp contrast between the complexity
of the full theory and relative simplicity of the final answers.   
It is a kind of disappointment to address, for instance,
the problem of linear conductivity of a double tunnel
barrier  that separates a normal metal and
a superconductor, to spend weeks and weeks calculating 
and to obtain that under very general assumptions
$G=G_N/\sqrt{2}$, $G_N$ being conductivity in the normal state.

The so-called "circuit theory of Andreev reflection" 
was born in an attempt
to enhance the applicability of the full theory. Initially it
was a purely pedagogical project.
I wanted to derive a primitive
model that illustrates the essential features of the full theory.
To my surprise, the model worked in a very efficient way.
So I could not resist temptation and published a scientific
article.\cite{Nazarov1}

Since then, the circuit theory has been reviewed \cite{CarloRev1,CarloRev2}
and improved. C. W. J. Beenakker, D. Esteve, M. Devoret, 
N. Argaman \cite{Argaman}
reported extensions and reformulations, 
published as well as unpublished. Very elaborated formulation
can be found in \cite{Gueron}. 
The author has extended the circuit theory to cover ballistic
point contacts that enabled me to make calculations reported in
\cite{Hartog}.

In the present work we present a novel formulation of
the circuit theory that incorporates most of the changes
proposed. This theory enables energy- and voltage-dependent
transport calculations in superconducting and normal structures.
An important extension reported here 
is the concept of an "arbitrary connector".
This allows for unified treatment of superconducting
junctions of arbitrary nature. 
This extension is also relevant for 
the full theory.

In fact, the resulting circuit theory looks very much like as a 
discrete version of the full theory. So that the full power of the
full theory can now be incorporated in a circuit calculation.
From the other hand, even a very sophisticated experimental
layout can be presented with a few circuit theory elements.
This essentially simplifies any practical calculation,
numerical as well as analytical one.

The outline of the paper is as follows. In Section \ref{old} we remind
the reader the approach of the old circuit theory and its limitations.
We formulate the requirements to a circuit theory that is to
overcome these limits. In Section \ref{disc} we provide a discrete
version of the full theory equations in the diffusive limit. 
We introduce an important concept of the "leakage current" to 
describe decoherence between electrons and holes.  
We extend the concepts of a node and a connector in the next section.
Section \ref{next} is devoted to the arbitrary connector. We
show there how a connector with known transmission
eigenvalues shall be incorporated into the circuit theory.
In section \ref{rules} we accomplish the formulation
by giving  the set of the rules of the new circuit theory
and provide a simple example in Section \ref{exem}.
Section \ref{numeric} is devoted to the numerical algorithm that 
seem to suit most the structure of the theory. We discuss its
practical implementations.

\section{Circuit theory: old and new}\label{old}
Physics of electric conduction in a system that consists of
superconductors and normal metals, with optional tunnel
junctions,
can be adequately treated with non-equilibrium superconductivity equations.
Those are written for Keldysh Green's functions .
\cite{LarkOvch1,LarkOvch2}
Those are $4 \times 4$ Nambu-Keldysh matrices
matrices depending in a stationary case on space coordinates and energy.
They can be subdivided onto $2 \times 2$ Nambu matrices made up from
usual and anomalous Green functions (we use 'check' for
$4 \times 4$ and 'hat' for $2 \times 2$ matrices):
\begin{equation}
\check G(x,x')=\left(
\matrix{\hat G^R(x,x') & \hat G(x,x') \cr 0 &\hat G^A(x,x') }
\right);
\hat G^{A}(x,x')=\left(
\matrix{  G^A(x,x') & F^A(x,x') \cr -F^{+A}(x,x') &  -G^{+A}(x',x) }
\right).
\end{equation}
and similar for $\hat G^R$ and $\hat G$.
Semiclassical and diffusive appoximations
allow to obtain equations for Green functions in coinciding points.
We define $\check G(x)= i \nu/\pi \check G(x,x)$,
\begin{equation}
\check G=\left(\begin{array}{c} \hat R \ \ \hat K \\ 0 \ \ \hat A \end{array}\right);
\end{equation}
Here advanced and retarded functions determine the characteristics of
energy spectrum of the system in a given point whereas $\hat K$ sets
particle distribution over these energy states and thus directly related
to the electric current and other physical quantities. It is assumed
that the size of the system at least in a transport direction greatly
exceeds Fermi wavelength and elastic mean free path, this makes
diffusive approximation sensible. In this approximation, non-equilibrium
superconductivity equations resemble a standard diffusion equation
and read \cite{LarkOvch1,LarkOvch2,Klapwijk}
\begin{eqnarray}
\frac{\partial}{\partial x^{\alpha}} \left( {\cal D}(x) \check G
\frac{\partial}{\partial x^{\alpha}} \check G \right) - i[\check H,\check G] =0 ;\\
\label{general}
\check H= \epsilon \hat \sigma_z+ i \hat \sigma_x {\rm Re}(\Delta(x))
+i \hat \sigma_y {\rm Im}(\Delta(x)).
\end{eqnarray}
Here $\Delta$ is superconducting pair potential, ${\cal D}$ stands for diffusivity.  
A pseudounitary condition
holds for the matrices $\check G(\epsilon)$: $\check G^2= \check 1$.
We introduce here a density of  a matrix current defined by
\begin{equation}
\check j^{\alpha}(x)=
\sigma(x) \check G \frac{\partial}{\partial x^{\alpha}}\check G,
\label{current}
\end{equation}
$\sigma$ being specific conductivity in the normal state, so that
the equation \ref{general} can be presented as a conservation law
for this current
\begin{equation}
\frac{\partial}{\partial x^{\alpha}}\check j^{\alpha}(x)-i e^2 \nu [\check H,\check G]=0;
\label{conservation}
\end{equation}

The matrix current density (\ref{current}) can be related to
the electric current density by means of
\begin{equation}
j^{\alpha}_{el}(x) = \frac{1}{4e} \int d\epsilon {\rm Tr} \left[
\sigma_z \hat j^K(\epsilon,x) \right]
\label{elcurrent}
\end{equation}  

If there are tunnel interfaces in the structure, $\check G$ are in
general different on both sides of the interface.
Eq. \ref{general} shall be supplemented with the boundary conditions 
at the interface. These conditions
have been derived in \cite{Kuprianov} and can be written in a
comprehensive form as
\begin{equation}
N^{\alpha} \check j^{\alpha}(x) = \frac{g(x)}{2} [\check G_1(x),\check G_2(x)],
\label{boundary}
\end{equation}
$N^{\alpha}$ being a vector normal to the interface at a point $x$,
$g(x)$ being the conductance of the interface per unit area at the
same point, $1,2$ refer to different sides of the interface.
These equations shall be also fulfilled by the boundary conditions
"at infinity", otherwise their solution is not  unique.
Since we are talking about the transport, there must be at least
two "infinities" that correspond to source and drain. 
In general, the structure can be connected to many bulk electrodes,
either normal or superconducting, so it can have many "infinities".
Such bulk electrode we will call "terminal".  
In each terminal $\check G$ shall assume its equilibrium
value corresponding to the voltage at which the terminal is biased.

With all these additions, Eq. \ref{general} provides a solid
framework to calculate electric properties of superconducting
structures.

The circuit theory of Andreev conductance \cite{Nazarov1}
has grown from the fact that, under certain conditions, it is plausible
to omit  the second term in Eq. \ref{conservation}. So that the
matrix current is conserved exactly. 

Now we spell limitations
under which it is a true thing to do so. The first term is of the order
of ${\cal D}/L^2$, $L$ being the system size, or more precisely, the size
at which the resistance of the structure is being formed. In the normal metal
$\Delta \equiv 0$ and the right term is of the order of $\varepsilon$.
Thus we need a sufficiently small system: $ L << \sqrt{{\cal D}/\epsilon}$. 
It implies that:
a) the temperature is low enough: $T \ll {\cal D}/L^2, T \ll \Delta$.
b) the voltage is low enough: $ eV \ll {\cal D}/L^2, V \ll \Delta$.
The last limitation below is given by the fact that we use 
stationary equation.
If there had been several superconducting terminals in the structure 
biased
by different voltages, it would have given rise to non-stationary
Josephson-like effect that would make the Green functions to depend
on two energies.
So that, c) all sureconducting terminals are at the same voltage. Let us
set this voltage to zero.

The key idea of any circuit theory is to get from continuous
conservation law of the type (\ref{conservation}) to its
discrete version. The structure is presented as a set of {\it nodes}
those are  connected pairwise by means of {\it connectors}.
The conservation law is presented as a set of
second Kirchhoff rules for each node expect terminal ones,
\begin{equation}
0=\sum_k \check I_{ik},
\label{Krules}
\end{equation}
where the summation goes over all the nodes $k$ connected to
the node $i$. The current $\check I_{ik}= -\check I_{ki}$
depends on the states of the nodes $i,k$ and on the connector.
Therefore, the Eqs. \ref{Krules} determine the state of
the nodes at a given state of the terminals.
In a normal terminal biased at voltage $V$
$\check G$ is given by
\begin{equation}
\hat R = - \hat A = \sigma_z; \ \ \hat K = 2 
\left(\matrix{ \tanh \frac{\epsilon +eV}{2T} & 0 
\cr 0 & -\tanh \frac{\epsilon -eV}{2T}} \right) . 
\label{normterm}
\end{equation}
In a superconducting terminal,
\begin{equation}
-h \hat A= h^{*} \hat R =\hat H; \ \ \hat K= (\hat R - \hat A)\tanh \frac{\epsilon }{2T}  
\label{supterm}
\end{equation}
where $h=\sqrt{(\epsilon-i \delta)^2 - \vert \Delta \vert^2}$.
In a common electric circuit theory the state of the node
is characterized by its electrostatic potential. The current
is a scalar given by $I_{ik}= g_{ik}(\phi_i-\phi_k)$.
In the circuit theory \cite{Nazarov1} the state of the
node is in principle characterized by the full $\check G(\epsilon)$
and the current retains matrix structure.
However the symmetries of $\check G$ specific for 
$\epsilon \approx 0$ allow for significant simplifications.
Energy dependence of $\hat A,\hat R$ can be disregarded and
$\hat K$ can be integrated over the energy. The Kirchhoff
equations can be separated on two parts. First part are equations
for vector "spectral currents" to determine 
"spectral vectors" $\hat A,\hat R$ in each node.
Second part are equations for usual currents to determine
non-equilibrium chemical potentials in each node and, finally,
Andreev conductance. The resistance of the tunnel connectors
is remormalized, this renormalization is given by
scalar product of two "spectral vectors" on the sides
of the connector. 

The old circuit theory is good to obtain simple answers.
If the analysis of experimental data in real structures is meant,
the conditions a) and b) appear to be very restrictive.
Beside the fact that the temperature can not be made
arbitrary low, the voltage and temperature dependence
of the transport can not be accessed in the framework
of the theory. \cite{Nazarov1}

We will not explain here the rules and details of the old circuit theory:
those can be found in \cite{Nazarov1,CarloRev1}. Our goal
is to formulate a circuit theory that is free form $\epsilon \approx 0$
limitation. We can now spell the requirements to the structure
of this theory.
First, in order to comply with the full theory,
the state of the node shall be characterized by a full
$\check G$ matrix. 
Second, since energy dependence of the
Green functions has to be taken into account, the 
equations (\ref{Krules}) shall be written and solved 
separately in each energy slice. By doing so we disregard
possible inelastic scattering, which is usually not important.
Physical values like electric current are then obtained
by integration over $\epsilon$.
The third requirement stems from the recent technological
achievements. It is possible now to incorporate (ballistic)
point contacts into mesoscopic superconducting structures.
So we require that the theory shall describe
such contacts and, in general, any arbitrary constrictions
and connections.

Below we derive the rules of this theory and discuss their
implementations.

\section{Discretization of duffusive conductor}\label{disc}

Let us begin with formulation of the discrete version
of Eq. \ref{general}. Instead of continuous space $x^\alpha$
we take a connected discrete set of $x^{\alpha}_i$ such that
$\check G(x^{\alpha}_i)$ in neighboring points of the set are
close to each other. We associate a resistor
with each connection in the set in such a way that it simulates
continuous resistivity of the system. Let us show how to choose
such resistors for square lattice of $x^{\alpha}_i$ with periods
$a^{\alpha}_{1,2}$(
$a^{\alpha}_1 a^{\alpha}_1 = a^{\alpha}_2 a^{\alpha}_2= a^2,
a^{\alpha}_1 a^{\alpha}_2=0$) that approximates homogeneous film
with sheet conductivity $\sigma$. Let us expand $\check G$ in the
vicinity of the node $i$: $\check G(x) = \check G(x_i) +
\check \Xi^\alpha (x^\alpha-x^{\alpha}_i)$, $\Xi^\alpha a <<G(x_i)$. 
 From Eq. \ref{current} we obtain the continuous matrix current density
\begin{equation}
\check j^{\alpha}= \sigma \check G(x_i) \check \Xi^\alpha
\label{continu}
\end{equation} 
In the network, the current density
is given by
\begin{equation}
\check j^{\alpha} = a^{-2} \sum_{k} \check I_{ik} (x_k^{\alpha}-x_i^{\alpha})
\end{equation}
$k$ being nodes neighboring $i$.
We see that if we choose 
\begin{equation}
I_{ik} = \frac{g_{ik}}{2} [\check G_i,\check G_k],
\label{TJ}
\end{equation}
$g_{ik} = \sigma a$ we reproduce Eq. \ref{continu}. To prove this
we make use of the fact that $[\check \Xi^{\alpha},\check G_i]_{+}=0$,
the latter follows from $\check G^2=1$. Now we can rewrite
Eq. \ref{conservation} as
\begin{equation}
0=\sum_k \check I_{ik} + \check I_{leakage};
\label{with_leakage}
\end{equation} 
where we introduce "leakage" current 
\begin{equation}
\check I_{leakage} = -i e^2 \nu V_i [\check G_i,\check H],
\label{leakage}
\end{equation}
$V_i$ being volume associated with the node $i$. The term
"leakage" comes from the fact that the Eq. has a formal
similarity with the equations of the electric circuit theory
that take into account the charge leakage to the ground from each 
node. 
There is no net leakage of the electric current due to
(\ref{leakage}) since the corresponding term appears to be zero. 
However, the matrix current (\ref{leakage}) 
describes two processes that may be viewed as a sort of leakage.
Namely, the terms proportional to $\epsilon$ describe decoherence
between electrons and holes, that is, the fact that the electrons
and holes at the same energy difference $\epsilon$ from Fermi
surface have slightly mismatching wavevectors. So we have
"leakage of coherence". The terms proportional to $\Delta$ are
responsible for the conversion between quasiparticles and cooper pairs
that form the superconducting condensate. This is a leakage of
quasiparticles.

We note that, by virtue of Eq. \ref{boundary}, the Kirchhoff rules
(\ref{TJ}), (\ref{with_leakage}) are also suitable for  systems
containing  tunnel junctions. Just some connections of the
resistive network shall be replaced with (specific) conductances
of the tunnel barriers.

With increasing fineness of the node set we obviously converge
to continuous limit and obtain more and more accurate
agreement with the results of the full theory. A practical
question is how fine should this mesh be in a realistic calculation.
A short answer is that the nodes shall be close than
than the coherence length $\xi= \sqrt{{\cal D} \max(\Delta,\epsilon)}$.
In practice, the mesh describing a diffusive wire
should contain about a dozen nodes even at
$\xi$ that exceeds the system size. This drawback is compensated
in the limit of short $\xi$, since in this case the interesting
behaviour occurs merely at the edge of the system and it makes
no sense to keep fine mesh through the whole sample. It
turns out that the mesh of $40$ nodes is sufficient
to describe transport properties of the diffusive wire
this the accuracy at least $10^{-4}$ in the whole energy range.

\section{More about nodes and connectors} \label{moree}

As we saw above, the essential requirements to a node
are i. its state is characterized by $\check G$, 
so that Green functions must be isotropic on the scale on the
node size, ii. the size of the node is less than coherence
length.

This allows us to treat as nodes  parts of the 
system where the transport is not or is not entirely
diffusive. Let us consider, for instance, the well-known
model of ballistic cavity.\cite{Cavity} Although the transport in the cavity
is the ballistic one, one can actually regard the cavity
as a node connected to reservoirs by means of ballistic point
contacts. It is the chaotic character of ballistic transport
that makes Green function isotropic. The condition of good
isotropization is fulfilled provided the diameter of the
cavity greatly exceeds the typical size of the point contacts,
or, in other terms, the conductance of the contacts must
be much less than the Sharvin estimation of the cavity conductance.
The cavity must be smaller than the coherence length in the
ballistic limit. The leakage current from the cavity
is still given by Eq. \ref{leakage}: it does not know which mechanism
has provided the isotropization of the Green function, the cavity
could be diffusive as well.

This makes it relevant to investigate such connectors between
the nodes as ballistic point contacts. Previous experience with
superconducting constrictions shows a significant difference
between tunnel junctions, diffusive and ballistic constrictions.
This difference stems from the fact that the constrictions 
of the same conductance have different distribution of transmission
eigenvalues.

This implies that the expression (\ref{TJ}) is not valid for the
arbitrary connector. Below we derive the expression for the
matrix current in terms of transmission eigenvalues.

\section{Arbitrary connector}\label{next}
The derivation presented below makes substantial use
of the approach \cite{Zaitsev} and the general scattering
theory outlined in \cite{Stone}. The goal is to express
the matrix current in the junction between two nodes
in terms of  isotropic Green functions in these nodes
and transmission eigenvalues that characterize the
scattering in the junction.
We subdivide the system onto five zones: two isotropization
zones on both sides of the junction, two ballistic zones
and scattering zone. We assume that the size of all zones
in transport direction is much shorter than the coherence
length. Under this condition, the matrix current conserves
and can be easily evaluated in ballistic zone where Green function
equations are simple. The same condition allows us to disregard
energy dependence of the scattering amplitudes. 

Following to \cite{Stone}, we discretize the transverse motion
of electrons.
We introduce a number of transport channels labeled by
$n$ and let $z$ to be  along the transport direction. The scattering
takes place  near $z=0$. Let us first consider the 
left side of the junction, $z<0$.
Similar to the approach of the ref. \cite{Zaitsev}
we present the exact Green function $\check G_{nm}(z,z')$
in the following form
\begin{equation}
\check G_{nm}(z,z')= \sum_{\sigma,\sigma'=\pm 1} 
\exp(i\sigma p_n -i \sigma' p_m) \check G^{\sigma\sigma'}_{nm}(z,z'),
\end{equation}
 $p_n$ being Fermi momentum in the $n$-th channel.
The functions $\check G^{\sigma\sigma'}{nm}$ are varying
smoothly at the scale of $1/p_n$ and obey the following
semiclassical equations:
\begin{eqnarray}
({\sigma v_n \frac{\partial}{\partial z} + \check {\cal H}}) 
\check G^{\sigma\sigma'}_{nm}=0 \\
\check G^{\sigma\sigma'}_{nm}(\sigma' v_n \frac{\partial}{\partial z'} - \check {\cal H})=0. 
\label{derivatives}
\end{eqnarray}
$\check {\cal H}$ being the effective Hamiltonian that comprises $\check H$
and the self energy part that describes isotropization scattering.
These functions are discontinuous at $z=z'$, their values for $z>z'$ and
$z<z'$ being matched with the aid of the following condition
\cite{Zaitsev}
\begin{equation}
\check G^{\sigma\sigma'}_{nm}(z+0,z)-\check G^{\sigma\sigma'}_{nm}(z-0,z)=
i {\check 1} \sigma\delta_{nm} \delta_{\sigma\sigma'} 
\end{equation}

It is convenient to work with matrices defined in an extended basis with
comprises Nambu-Keldysh indices, channels, and $\sigma$= direction of the
mode.
We will use tilde accent for matrices in this basis, for example,$\tilde A$. We will
also denote with $\check A$ the matrices that are diagonal in mode indices
and with $\bar A$ those diagonal in Nambu-Keldysh indices.
To characterize the Green functions $\check G^{\sigma\sigma'}_{nm}$ we introduce
the matrix $\tilde g$ such that
\begin{equation}
2i \check G^{\sigma\sigma'}_{nm}(z,z') = 
{\check g}^{\sigma\sigma'}_{nm}(z,z')/\sqrt{v_n v_m} + \sigma
\delta_{\sigma\sigma'}{\rm sign}(z-z')/|v_n|
\end{equation}  
The matrix $\tilde g$ is continuous at $z=z'$. The factors $v_n$ are
chosen in such a way that $\tilde g$ is composed of the scattering
wave functions that are normalized per unit flux. This makes it easy
to apply the general scattering formalism of Ref. \cite{Stone}.
Following \cite{Stone} we introduce a transfer matrix that relates
wave functions on the right side ($\psi^{(2)}$) and those on the left side
($\psi^{(1)}$)
of the scatterer,
\begin{equation}
\psi^{(2)}_{m,\sigma}= \sum_{n,\sigma'}
M_{m,\sigma;n,\sigma'}\psi^{(1)}_{n,\sigma'}
\end{equation} 
Since $\tilde g$ transforms as a product of wave functions, its values
at $z,z'>0$ and $z,z'<0$ are related by
\begin{equation}
\tilde g_2 = \bar M^{\dagger} \tilde g_1 \bar M.
\label{scattering}
\end{equation}
The transfer matrix $M$ obeys flux conservation ,
\begin{equation}
\bar M \bar \Sigma^z \bar M^{\dagger} = \bar \Sigma^z,
\label{conservation_p}
\end{equation}
where $\bar \Sigma^z_{\sigma,m;\sigma,n} \equiv \sigma \delta_{mn}
 \delta_{\sigma\sigma'}$. 
It follows from (\ref{conservation_p}) and (\ref{scattering})  that the quantity we are after, 
the matrix current,
can be expressed in terms of $\tilde g$ on either side:
\begin{equation}
\check I= \frac{e^2}{\pi}{\rm Tr}_{m,\sigma} 
\left[\bar \Sigma^z \tilde g_1 \right]
=\frac{e^2}{\pi}
{\rm Tr}_{m,\sigma} \left[ \bar \Sigma^z \tilde g_2 \right].
\label{current_p}
\end{equation} 
To find the matrix current, we have to evaluate $\tilde g_{1(2)}$.
To do so, we shall consider the behavior of $\tilde g$ in the
isotropization zone on both sides of the scatterer. 
Let us first consider the left side of the scatterer and let us
assume the simplest model of the isotropization: scattering
on the point defects. Then in the
isotropization zone we may approximate $\check {\cal H} = \check G_1/2\tau_{imp}$.
Since $\check {\cal H}$ does not depend on $z$ the equations \ref{derivatives}
can be readily solved in the most general form and we can express
$\tilde g(z,z')$ in the isotropic zone in terms of its value $g_1$
in the ballistic zone,
\begin{equation}
\check G^{\sigma\sigma'}_{nm}(z,z') = \tilde P(z) ( \tilde g_1 +\bar
{\rm sign}(z-z') \Sigma^z) \tilde P(-z');
\end{equation}
where $\tilde P$ is diagonal over channel index,
\begin{equation}
\tilde P(z)= \frac{\delta_{nm}}{2} ( 
\exp(\frac{z}{2 v_n \tau_{imp}}) ( \tilde 1 - \bar \Sigma^z \check G_1)+
\exp(-\frac{z}{2 v_n \tau_{imp}}) (\tilde 1 + \bar \Sigma^z \check G_1))
\end{equation}
To derive that we use that $\check G_1^2=\check 1$. 
To assure that $\check G^{\sigma\sigma'}_{nm}$ does not grow with
decreasing $z$, we shall require
\begin{equation}
(\bar \Sigma^z + \check G_1)( \bar \Sigma^z - \tilde g_1)=0;
\label{asympt_een}
\end{equation} 
\begin{equation}
( \bar \Sigma^z + \tilde g_1)(\bar \Sigma^z - \check G_1)=0.
\label{asympt_twee}
\end{equation}
Similar consideration for the right side of the scatterer yields
\begin{equation}
(\bar \Sigma^z - \check G_2)( \bar \Sigma^z + \tilde g_2)=0;
\label{asympt_drie}
\end{equation}
\begin{equation}
( \bar \Sigma^z - \tilde g_2)(\bar \Sigma^z + \check G_2)=0.
\label{asympt_vier}
\end{equation}

We see that the above conditions do not contain any information about
how the isotropization actually occur. This  suggests that
those are of universal nature and do not depend on isotropization
mechanism.

Conditions (\ref{asympt_een})-(\ref{asympt_vier}) together with Eq. \ref{scattering} 
completely determine $\tilde g_{1(2)}$, Green function in the ballistic region.
To find an explicit expression, we multiply Eq. \ref{asympt_een} by $\bar M^{\dagger}$
from the left and by $\bar M$ from the right. Taking into account Eqs. 
\ref{scattering} and \ref{conservation_p} we find that
\begin{equation}
\bar Q + \check G_1 \bar \Sigma^z - ( \bar Q \Sigma^z + \check G_1) \tilde g_2 =0.
\label{intermediate}
\end{equation}  
Following \cite{Stone}, we have introduced here the matrix 
$\bar Q \equiv \bar M^{\dagger} \bar M$, 
$\bar Q^{-1} = \bar \Sigma^z \bar Q \bar \Sigma^z$.
Now we multiply Eq. \ref{asympt_drie} by $\bar Q$ from the left and add
it to Eq. \ref{intermediate}. We derive $\tilde g_2$ from the resulting 
equation,
\begin{equation}
\tilde g_2 = \frac{\tilde 1}{\bar Q \check G_2 + \check G_1} 
(2 \bar Q +(\check G_1 - \bar Q \check G_2) \bar \Sigma^z)).
\label{g2}
\end{equation}
After some algebra it is possible to check that expression
(\ref{g2}) satisfies the conditions (\ref{asympt_een})-(\ref{asympt_vier}) provided $\check G_1^2= \check G_2^2
=\check 1$, as it should be. Besides, $\tilde g_{1(2)}^2 = \tilde 1$.
The equation \ref{g2} presents an important step in the derivation.

To evaluate the matrix current we find $\tilde g_2$ in the basis
composed of eigenvectors of $\bar Q$ and Nambu-Keldysh indices. In this basis
$\tilde g_2$ takes a blockdiagonal form. It appears \cite{Stone} that
for each eigenvector $\vec{c}_n$ with eigenvalue $q_n >1$ the vector 
$\bar \Sigma^z \vec{c}_n$ is also an eigenvector of $\bar Q$ with
eigenvalue $q_n^{-1}$.
These two eigenvectors form a block. Within the block
\begin{equation}
\bar Q = \left( \matrix{ q_n& 0\cr
0& q_n^{-1}\cr} \right); \ \ \ \bar \Sigma^z = \left( \matrix{ 0& 1\cr
1& 0\cr} \right).
\end{equation}
This allows us to find $\tilde g_2$ within the block:
\begin{equation}
\tilde g_2 =  \left( \matrix{ \frac{2 q_n}{q_n \check G_2 + \check G_1}&
\frac{1}{q_n \check G_2 + \check G_1} (\check G_1- q \check G_2)\cr
\frac{1}{q_n \check G_1 + \check G_2} (q_n\check G_1-  \check G_2)&
\frac{2 }{q_n \check G_1 + \check G_1}\cr} \right).
\end{equation}
Now we can directly apply Eq. \ref{current_p} to evaluate the matrix
current. Finally we obtain
\begin{equation}
\check I = \frac{e^2}{\pi} \sum_n 2 T_n \frac{\check G_2 \check G_1 - \check G_1 \check G_2}
{4+T_n(\check G_2 \check G_1+ \check G_1 \check G_2-2)}
\label{final}
\label{arb_current}
\end{equation}
where we have switched to more conventional representation of $q_n$ 
in terms of transmissions $T_n$, eigenvalues of the transmission matrix
square.\cite{CarloRev2}

Taking  its practical use apart, the relation (\ref{final}) is probably
the most concise way to express the accumulated knowledge
about superconducting constrictions. We note that the electric
current in the constriction is given by the Keldysh part
of (\ref{final}),
\begin{equation}
I_{el}= \frac{1}{4e} {\rm Tr} \left[\sigma_z \hat I^{K}\right].
\end{equation} 
If we set one of $\check G$ to the values of a normal terminal (\ref{normterm})
and another one to the values of superconducting terminal (\ref{supterm})
we obtain i. Beenakker's relation for Andreev conductance \cite{CarloRev2}
in the limit $\epsilon \rightarrow 0$, ii. a simple generalization
of Blonder-Tinkham-Klapwijk formula \cite{Blonder} at $\epsilon$
of the order of $\Delta$. Setting both $\check G$ to superconducting
terminals of different phases give all known relations for the
Josephson current.  

Keldysh components of (\ref{final}) can be separated.
For retarded one we have
\begin{equation}
\hat I^R = \frac{e^2}{\pi}\sum_n \frac{T_n}{2+T_n ({\bf R}_1 \cdot {\bf R}_2 -1)} [\hat
R_2,\hat R_1]
\end{equation}
and similar for $A$. This suggest that we can characterize
the scatterer with a single scalar function $Z(x)$,
\begin{equation}
Z(x) \equiv \frac{e^2}{\pi}\sum_n \frac{T_n}{2+T_n 
(x-1)}.
\end{equation}
For any constriction $Z(1)=G_N/2$.

The expression for Keldysh component looks rather cumbersome: 
\begin{equation}
I^K= Z({\bf R}_1 \cdot {\bf R}_2) \hat R_1 \hat K_2+
\frac{Z({\bf R}_1 \cdot {\bf R}_2)-Z({\bf A}_1 \cdot {\bf A}_2)}
{2({\bf R}_1 \cdot {\bf R}_1-{\bf A}_1 \cdot {\bf A}_2)}
\left( 2 \hat K_2 \hat A_2 + \hat R_1 K_2 \hat A_1 \hat A_2
- \hat R_2 \hat R_1 \hat K_2 \hat A_1 \right) -
Z({\bf A}_1 \cdot{\bf A}_2) \hat K_2 \hat A_1 - \left( 1 \leftrightarrow 2
\right)
\label{IK}
\end{equation}
It is conventional to present $\hat K$ in the form
$\hat R \hat f - \hat f \hat A$, where $\hat f= f^{L} \hat 1+ f^T \hat \sigma_z$
presents two-component distribution function in the superconductor.
We note that the expression (\ref{IK}) has in fact only
two independent components, that corresponds to two components
of the distribution function. One can choose these components
to be $I_0={\rm Tr} \hat I^K$, $I_z={\rm Tr} \sigma_z \hat I^K$.
Then one can express these components in terms of $f^{T,L}$ on both
sides of the junction. Later one can keep
only these components in  the Kirchhoff rules (\ref{with_leakage}).
However, the corresponding expressions are so cumbersome
that we do not dare to recommend this way: it is simpler
to work with all four components of $\hat I^{K}$.

Albeit there is an important case where the presentation
mentioned helps a lot. If there is no mixing of normal and
superconducting current whithin the system, equations for
$I_0$ and $I_z$ separate, with $I_z$ depending on  $f^{T}$ only, 
\begin{equation}
2I_z(\epsilon) = G(\epsilon)(f_2^{T}(\epsilon)- f_1^{T}(\epsilon));
\end{equation}
where $G(\epsilon)$  reads
\begin{equation}
G(\epsilon)= 2 {\rm Re} \left( Z(\xi) \xi \right)+
2 {\rm Re} Z(\xi){\rm Re} \mu+
\frac{Z(\xi)-Z(\xi^*)}{2 {\rm Im} \xi} \left( \vert \mu \vert^2
+ \vert \xi \vert^2 + 2 {\rm Re} \mu{\rm Re} \xi
-(1-s_1)(1-s_2)\right)
\label{gen_renorm} 
\end{equation}
where 
\begin{equation}
\mu= {\bf A}^+_1 \cdot {\bf A}_2; \ \ \xi={\bf A}^+_1 \cdot {\bf A}^+_2;
\ \ s_{1,2} = {\bf A}^+_{1,2} \cdot {\bf A}_{1,2}.
\end{equation}
It is important that $G(\epsilon)$ 
can be viewed as energy-dependent effective
conductance of the junction. This follows from the fact
that the Kirchhoff rules that determine $f^T$ at any
given energy can be viewed as electric circuit theory
rules with energy-dependent conductances.
 
For tunnel junctions, Eq. \ref{gen_renorm} reduces to
\begin{equation}
G(\epsilon) =\frac{G_N}{8} {\rm Tr} 
\left[ (\hat A_1 +\hat A_1^+)(\hat A_2 +\hat A_2^+) \right]
\label{renorm}
\end{equation}

The "no mixing" condition is always satisfied if there is
only only superconducting terminal in the system. It
can be satisfied if the normal and superconducting
current are geometrically separated like in the experiment. 
\cite{Petrashov} 

\section{Rules}\label{rules}
We summarize the above derivation by giving the rules of the
novel circuit theory.

i. Define the circuit. That includes a proper choice
of the node mesh in the diffusive parts of the system,
determination of the type of the connectors in terms
of $Z(x)$. 

ii. Write down Kirchhoff rules (\ref{with_leakage}) with leakage current.
Information about connectors shall be used at this point.

iii. Setting terminals to given values (\ref{normterm},\ref{supterm}),
find from the Kirchhoff rules $\check G$
in each node for each energy.

iv. Find currents in the circuit with the aid of Eqs. 
\ref{IK},\ref{TJ},\ref{elcurrent} 

These rules are to apply in the most general case.
If "no mixing" condition is satisfied, Kirchhoff rules
have to be solved for $\hat R, \hat A$ only. The 
energy-dependent conductances of all connectors are then
given by Eq. \ref{gen_renorm}.

In the rest of this section we give a simple relation
that allows practical work with Eqs. \ref{with_leakage} 
and occasionally
gives analytical results. The point is that by virtue
of (\ref{arb_current}) $\check I_{ik}$ can be always presented
in the form of $\check I_{ik} = [\check G_i,\check L_{ik}]$,
$\check L$ being a matrix.
So that the second Kirchhoff rule for the node $i$
takes the commutator form
\begin{equation}
0=[\check G_{i}, \check{\cal G}_i].
\label{commutator}
\end{equation} 
In the simple case where all connectors are of tunnel nature
expression for $\check{\cal G_i}$ reads
\begin{equation}
{\cal G}_i = \sum_k g_{ik} \check G_k + e \nu V_i \check H 
\label{calG}
\end{equation}
and does not depend on $G_i$. In general, it depends on $G_i$
by virtue of Eq. \ref{arb_current}. In any case the Eq. \ref{commutator}
allows to express $\check G$ in terms of ${\cal G_i}$. We explicitly
write Keldysh components of these matrices:
\begin{equation}
\check G = \left(\matrix {\hat R & \hat K \cr 0 & \hat A \cr}\right);
\check {\cal G} = \left(\matrix {\hat {\cal R} & \hat {\cal K} \cr 0 & \hat {\cal A} \cr}\right);
\end{equation}
For diagonal blocks we find that
\begin{equation}
\hat A = \hat {\cal A}/{\bf a}; \  \ \hat R = \hat {\cal R}/{\bf r};
\label{forA}
\end{equation}
where ${\bf a} =\sqrt{{\rm Tr}({\cal A}^2)/2}$, ${\bf r}= {\bf a}^*$.
Substituting this into (\ref{commutator}) we obtain 
that 
\begin{equation}
\hat K = ({\bf a} +{\bf r})^{-1} 
(\hat {\cal K} - \hat R \hat {\cal K} \hat A).
\label{forK}
\end{equation} 
\section{Simple example}\label{exem}
To give a simple application of the circuit theory rules
we consider a system of two tunnel junctions
of the same resistance $R_T$ with normal metal in between
that separate a superconducting electrode and a normal
metal biased at voltage $V$. If we assume that the correlation
length is shorter than the size of the intermediate
normal metal and its resistance is negligible in comparison
with the resistances of the tunnel junctions we can regard
it as a node. This node is connected to two terminals.
With the aid of (\ref{supterm}),(\ref{normterm}),
(\ref{calG}),(\ref{forA})
we find the Green function in the node. To simplify the
formulas, we write it explicitly only for $\epsilon \ll \Delta$:
\begin{equation}
\hat A = \frac{1}{\sqrt{(1+i \epsilon \tau)^2+1}}\left( 
\matrix{1+ i \epsilon \tau & 1 \cr 1 & -1 - i \epsilon \tau\cr} \right);
\end{equation}
where  $\tau = e^2 \nu V_{node} R_T$ is a typical escape time
from the node.
Since there is no superconducting current in the system,
we do not have to write $\hat K$ explicitly. Instead,
we evaluate the energy-dependent resistances for
each junction with the aid of Eq. \ref{renorm}
\begin{equation}
R_1 =\frac{R_T}{ 
{\rm Re}\left(
1/\sqrt{(1+i \epsilon \tau)^2+1}
\right)
}; \\
R_1 =\frac{R_T}{ {\rm Re}\left((1+i \epsilon \tau)/\sqrt{(1+i \epsilon
\tau)^2+1}\right)}.
\end{equation}
The differential conductance $G(V)$ of the system is then given
by $1/(R_1+R_2)$ at $\epsilon=eV$ if $T=0$. Fig.1 presents plots
of $R_T/R_{1,2}, G(V)$.  
\section{Numerical implementation} \label{numeric}
We have developed a simple numerical algorithm that
suits the structure of the circuit theory proposed.
It is based on successive iterations of Eqs. \ref{forA},\ref{forK}
at a given energy. Initially, an approximation for $\check G$
in each node is stored in memory. Next step is calculation of
$\check {\cal G}$ for each node. Here we make use of information
about connections in the circuit. From Eqs.\ref{forA},\ref{forK}
we obtain then next approximation for $\check G_i$. This algorithm
is not expected to converge very fast. In fact, the calculation
time is proportional to $N_{nodes}^3$. However, the algorithm 
does not make use of any complicated parametrization of $\check G$, 
that speeds up the calculation. Also, Green functions usually
have to be found at many close energies and are calculated
consequently. In this case, one
can use the final result for $\check G$ at a certain energy
as a good initial approximation for the next energy. 
All this makes the algorithm fast enough to suit
the practical purposes.

\acknowledgements
I am very much obliged to B. J. van Wees and S. den Hartog
for their persistent interest in circuit theory methods,
that has inspired me to have this job done.
I am indebted to C. W. J. Beenakker, D. Esteve, M. Devoret, 
H. Pothier, M. Feigelman, G. E. W. Bauer, B. Z. Spivak,
N. Argaman and J. E. Mooij
for many instructive discussions.
This work is a part of the research programme of the "Stichting voor
Fundamenteel Onderzoek der Materie"~(FOM), and I acknowledge the financial
support from the "Nederlandse Organisatie voor Wetenschappelijk Onderzoek"
~(NWO).

\begin{figure}
\epsfxsize=15cm
\epsfbox {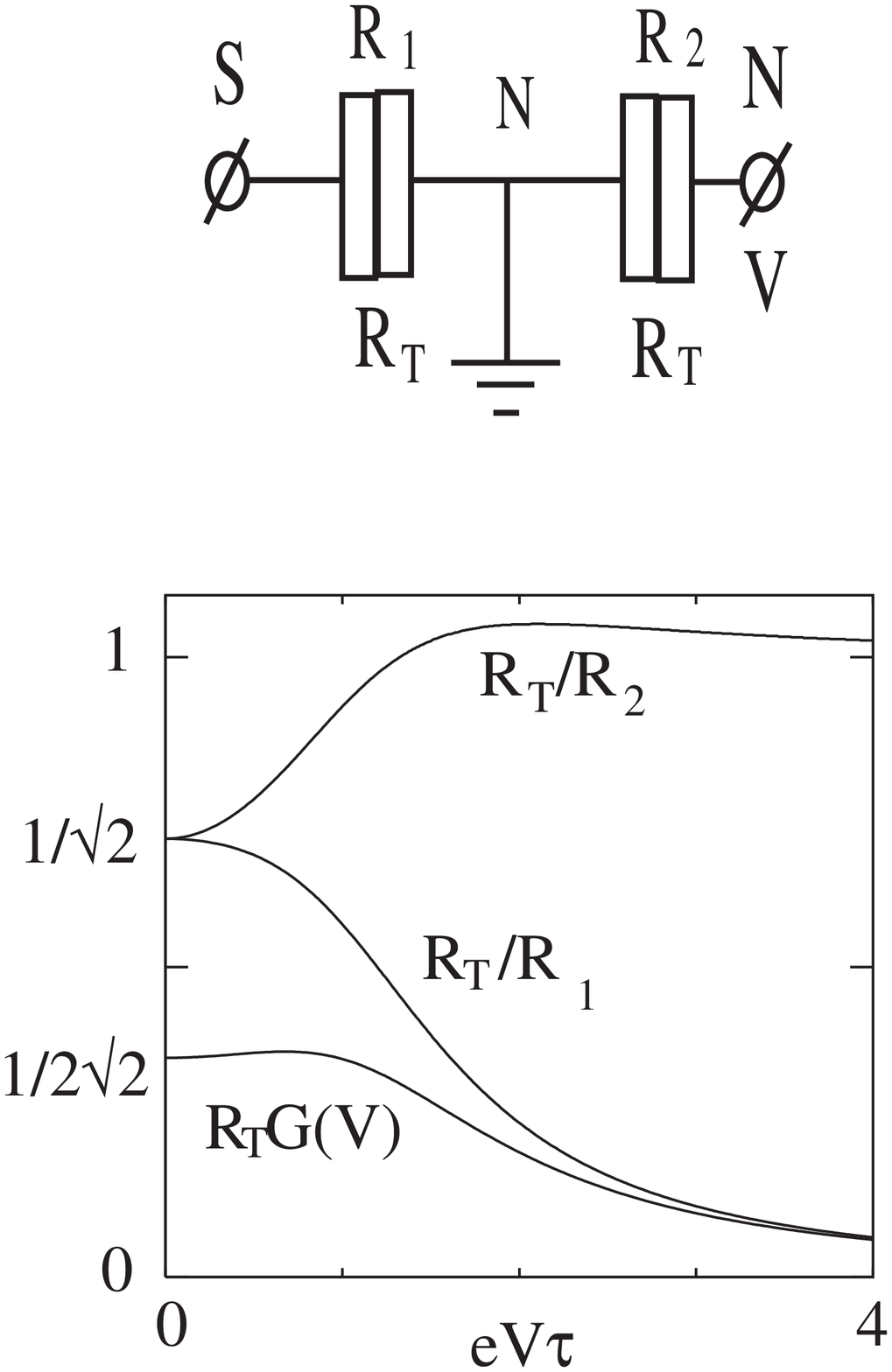}
\caption{
Double tunnel junction in circuit theory. Connection
to the ground represents the leakage current.
Effective conductances of both junctions along with
the net conductance are plotted versus voltage applied.
}
\label{fig1}
\end{figure}

\end{document}